# A Firefly-inspired Model for Deciphering the Alien

Cameron Brooks[1], Estelle Janin[1], Gage Siebert[1], Cole Mathis[2,4,5], Orit Peleg[3,6,7], Sara Imari Walker[1,2,3]*

[1] School of Earth and Space Exploration, Arizona State University, Tempe AZ USA

[2] Beyond Center for Fundamental Concepts in Science, Arizona State University, Tempe AZ USA

[3] Santa Fe Institute, Santa Fe, NM USA

[4] School of Complex Adaptive Systems, Arizona State University, Tempe AZ USA

[5] Biodesign Center for Biocomputing, Security and Society, Arizona State University, Tempe AZ USA

[6] BioFrontiers Institute, University of Colorado, Boulder, Boulder CO USA

[7] Departments of Computer Science, Physics, Applied Math, and Ecology and Evolutionary Biology, University of Colorado, Boulder, Boulder CO USA

*author for correspondence: sara.i.walker@asu.edu;

## Abstract

The Search for Extraterrestrial Intelligence (ETI) is, historically, a search for aliens like us, inspired by human-centric ideas of intelligence and technology. However, humans are not the only instance of an intelligent, communicating species on Earth —and thus not guide to how we might think about ETI. Here, we explore the potential for the study of non-human species to inform new approaches in SETI research, using firefly communication patterns as an illustrative example. Fireflies communicate their presence through evolved flash patterns distinct from complex visual backgrounds. Extraterrestrial signals may also be identifiable not by their complexity or decodable content, but by the structural properties of the signal, as currently being explored in efforts to decode communication in non-human species across our biosphere. We present a firefly-inspired model for detecting potential technosignatures within environments dominated by ordered astronomical phenomena, such as pulsars. Using pulsar data from the Australia Telescope National Facility, we generate simulated signals that exhibit evolved dissimilarity from the surrounding pulsar population. This approach shifts focus from anthropocentric assumptions about intelligence toward recognizing communication through its fundamental structural properties—specifically, evolutionarily optimized contrast with natural backgrounds. Our model demonstrates that alien signals need not be inherently complicated nor need we decipher their meaning to identify them; rather, signals might be distinguishable as products of selection. We discuss implications for broadening SETI methodologies, leveraging the diverse forms of intelligence found on Earth.

# Introduction

The Search for Extraterrestrial Intelligence (SETI) is often cast as a search for something familiar by targeting signals that mirror humanity's current stage of technological development. For instance, the earliest indicators of technological activity from our planet came from the invention of radio transmission by the human species. The significance of our own radio broadcasts, and the fact that radio can be transmitted readily over cosmic distances, led to these wavelengths becoming a prime target for early SETI efforts (Sullivan et al., 1978). However, in subsequent decades, Earth has become less radio loud, not more, especially due to the transition from analog-TV to cable TV and the internet (Saide et al., 2023). This example highlights a challenge associated with looking for short-lived human technologies as a window into alien technologies, suggesting new paradigmatic frameworks will be necessary to recognize intelligent alien signals in astronomical data.

Detecting extraterrestrial intelligence (ETI) is challenging because it requires imagining a space of potentially highly exotic intelligences, which may have evolved in a manner completely different to the development of technological life here on Earth. To identify evidence of technospheres beyond our own, it will be critical to develop less anthropocentric approaches and cast as wide a net as possible for the kind of life we might find. Indeed, one of the primary focuses of both the Search for Extraterrestrial Intelligence (SETI) and the Messaging to Extraterrestrial Intelligence (METI) has been the identification or construction of a signal that would be universally understood regardless of language, culture, physiology, or psychology. Historically important efforts like Drake's Project Ozma exemplified this approach by targeting the 21-cm emission line of neutral hydrogen, a frequency presumed to be universally recognizable to radio-capable civilizations (Drake & Sagan, 1997). Similarly, the Arecibo message used prime numbers as a structural framework for its interstellar transmission, aimed at facilitating message decryption by potential recipients in the M13 globular cluster (The Staff at the National Astronomy and Ionosphere Center, 1975). These early initiatives have inspired a diverse array of subsequent SETI/METI endeavors. The Voyager Golden Records, launched in 1977, represent a multifaceted attempt at interstellar communication, incorporating audio, images, and scientific data encoded on gold-plated phonograph records (Lomax & Sagan, 1977). More recently, the Breakthrough Listen initiative, started in 2015, uses advanced radio and optical telescopes to survey a million nearby stars and 100 nearby galaxies across a wide range of frequencies looking for technosignatures—notably narrowband radio lines and optical laser pulses (Worden et al., 2017).

Though ingenious, these methods are likely still too centered on the human experience and way of understanding the world. These should not be expected to be readily conceived of, or even identifiable, by an alien mind. Likewise, alien communication may be entirely unlike the kinds of communication we find across the human species and human technologies. Earth is home to a rich diversity of organisms that differ enough from humans to provide insight into a more generalizable approach to detecting communication. Efforts like the Earth Species Project[1] are already advancing this agenda—using machine learning to decode communication across taxa, curating large open datasets, and developing species-independent representations that recover structure, syntax, and, where possible, semantics without human labels. This kind of cross-species decoding offers a concrete roadmap for recognizing nonhuman signal systems and testing inference methods that do not assume human priors. In general, studying how Earth's diverse intelligences communicate could provide more universal insights into the ways aliens may communicate, enhancing our chances of confident detection of an extraterrestrial signal.

Here, our goal is to provide a first astronomical model of alien communication, inspired by a non-human species. We focus on fireflies as a case study of a non-human organism which can communicate using a binary encoding. During the mating season, fireflies produce periodic flash sequences that are species-specific and help them identify conspecifics (Lewis and Crastley, 2008). When multiple species of fireflies are present in the same geographical and temporal location during mating season, their species-specific flash patterns minimize the risk of predation while being maximally distinct from other firefly species (Stanger-Hall & Lloyd, 2015; Woods Jr. et al., 2007). We build on the firefly communication model of Nguyen, Huang, and Peleg (2022), which simulates the evolution of firefly flash sequences over several generations, to examine how such signaling might influence the detectability of a "firefly-like" ETI against an astrophysical background. We develop a toy model which generates an evolved signal that is maximally distinct from a background of highly ordered, naturally occurring signals generated by pulsars. The signal we evolve minimizes energy consumption, in the same way that firefly flash patterns maximize distinctiveness from other firefly species while minimizing predation risk.

---

[1] https://www.earthspecies.org/

We use pulsars as the relevant astrophysical background in this study, because they are widespread throughout the galaxy and produce highly ordered emissions at regular intervals. Their behavior closely resembles what we might expect of an extraterrestrial signal (Tarter, 2001). Indeed, when pulsars were originally discovered in 1967, the shortness of their pulses alongside their regular periodicity were so unlike any known natural phenomenon at the time, that the possibility that the source of these signals was of alien design was considered (Penny, 2013). Although it was eventually determined that pulsars are natural phenomena, their similarity to our expectations of an intelligently designed signal makes them excellent baselines with which to compare to an evolved ETI signaling pattern using an example of historical notoriety. In addition, pulsars provide an apt analogue to firefly flash behavior: both produce discrete, regularly timed flashes that can be treated simply as sequences of on/off events. Because pulsar signals are also bright and readily detectable across interstellar distances, they offer a practical and observable backdrop against which to search for an extraterrestrial signal exhibiting evolutionary optimization for distinctness relative to its pulsar neighborhood. In other words, we are interested in cases where an observed signal might indicate the existence of an alien intelligence, modeled after a species other than humans. Such a signal would likely warrant further investigation and could be a strong candidate for having an extraterrestrial origin.

## Methods

During mating season, male fireflies utilize species-specific on/off light sequences that repeat periodically to signal their presence and availability to females. To aid in species recognition amongst the intense visual clutter present when numerous species and individuals swarm, many species have evolved to have flash sequences that minimize their similarity to other species of firefly present in the same environment (Lewis & Cratsley, 2008). It is unknown whether most ETI will want to communicate their presence and distinct identity in a manner akin to how these fireflies communicate their species and individual identity. However, this evolutionary problem faced by fireflies in densely packed swarming environments provides an opportunity to study how an intelligent species might evolve signals to identify its presence against a visually noisy astrophysical environment, using a non-human species as the model system of interest.

For fireflies, it is the punctual temporal pattern of the flash sequence—rather than sheer brightness—that most strongly influences mate attraction (Copeland & Moiseff, 2015). Thus, selection on flash structure reflects two pressures: lowering predation risk and energy use to increase fitness (Nguyen et al., 2022), and preserving species-specific distinctiveness so females recognize conspecific males. These competing demands create an optimization trade-off shaping signal evolution. Nguyen et al. (2022) modeled this by minimizing a two-part cost function: (i) a similarity metric—the sequence's average similarity to other species—and (ii) a predation-risk term. The similarity between two sequences is calculated as the length of the maximal possible overlap between sequences (see SI Section I). The predation risk value is calculated as the fraction of the sequence's length spent flashing or "on" (see SI Section I). Later in the ETI firefly model the predation risk will be equated with an energy cost. Weights were assigned to the similarity value and predation risk value to represent the trade-off between the two and can be adjusted to change the structure of the sequence produced (see SI Section I).

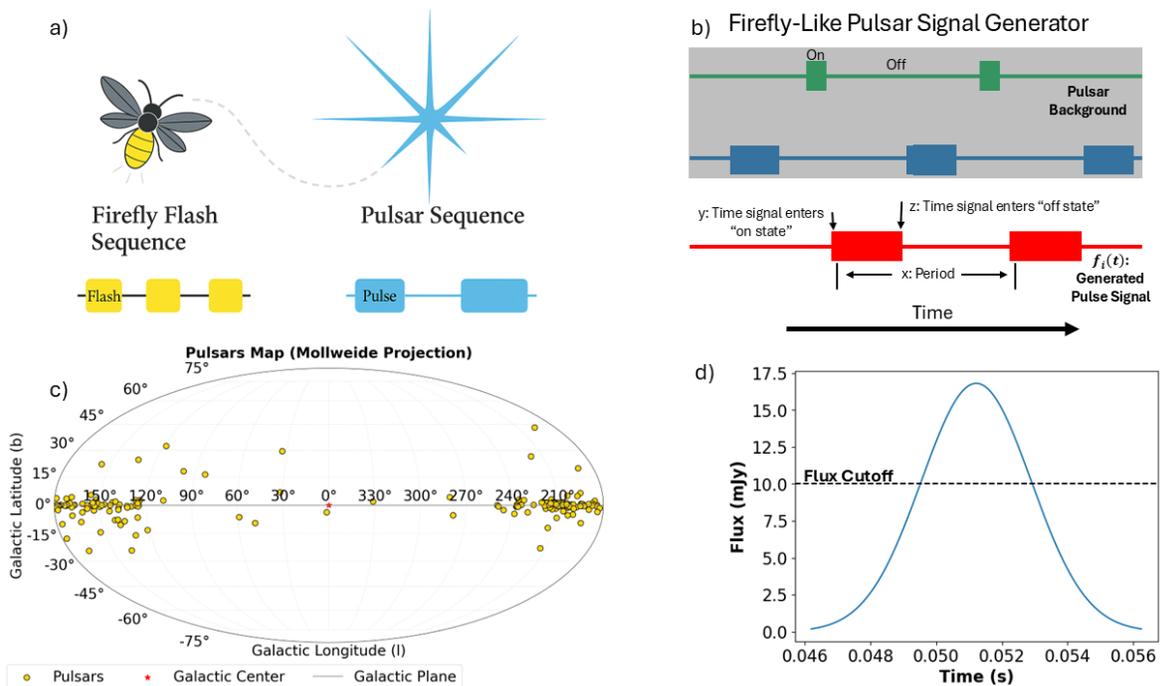
5

**Figure 1**: **a)** Illustration of a firefly flash sequence and a pulsar pulse sequence, which serves as the basis of the model discussed in this paper. **b)** Conceptualization of how the model creates a maximally distinct signal by minimizing the overlap of "on" and "off" sections between an artificial signal and the existing background signals. **c)** A Mollweide Projection displaying the galactic longitude and latitude of the 158 pulsars defining the pulsar background in the test case of this study. Background pulsars all have with mean flux densities greater than 1 mJy at 1400 MHz and are located within 5 kpc of the Earth. **d)** Representation pulsar pulse profile. The flux cutoff of 1 mJy determines which portion of the signal is considered "on" (above the cutoff) and "off" (below the cutoff).

We use data of 3724 pulsars from the Australia National Telescope Facility (ATNF) database (Manchester et al., 2005) (see SI Section II). This data is filtered depending on several parameters, such as radio frequency of interest, minimum mean flux density of the pulsar at a given radio frequency, galactic position, and size of the search area (see SI Section II). We binned the pulse profiles of these pulsars into "on" and "off" states reminiscent of the firefly flashes "on" and "off" states, using the user assigned mean flux density as a cutoff, where "on" represents a flux greater than the cutoff. Note that both firefly and pulsar flashes are analog, continuous signals binned to binary configurations, which inevitably discards potentially informative substructure (e.g., ramp-up/ramp-down shape and duration), whose detectability and encoded information would merit dedicated analysis. **Fig. 1** presents an overview of the study's conceptual framework, illustrating the firefly-pulsar analogy, the signal-generation model, the spatial distribution of the background pulsars, and a representative pulse profile.

The difference in magnitudes of the periods and absolute duty cycles of pulsars (i.e., the fraction of the period spent pulsing—in the "on" state) in this database can be very large, such that one pulsar could have a period of several hundred milliseconds and an absolute duty cycle of tens of milliseconds while another pulsar has a period of a few milliseconds and absolute duty cycle of a hundredth of a millisecond. Attempting to create a binary string length that would allow the features of both ends of this difference in magnitude to be represented and compared can result in each pulsar being tens of thousands of bits long. The model can potentially compare hundreds to thousands of pulsars in each iteration of the artificial signal being generated to find common substrings, but doing so by comparing signals bit by bit quickly becomes resource intensive. Furthermore, since there is no way to guarantee that the pulsar signals and the artificial signal will begin pulsing at the same time, the amount of time that both signals spend continuously in the same state will differ depending on when the overlap is observed and for how long one observes it. We attribute this phenomenon to two related factors: phase shift and phase offset. A phase shift happens when a periodic signal is displaced in time, causing it to pulse earlier or later within its cycle relative to another signal. A phase offset refers to the signal's starting point within its cycle—for example, a signal might begin right at the start of a pulse, or it might start halfway through the pulse. To ensure that the model accounts for phase shift and phase offsets, the model would have to either sufficiently extend the length of the binary string or generate an appropriate number of permutations of every single signal to capture all potential common substring lengths—which results in a high computational burden. To address this, instead of using binary strings of the same length like the Firefly Vocabulary Algorithm in Nguyen et al. (2022), we define a pulse signal, whether generated by a pulsar, a firefly, or artificially, as a sequence, $f_i(t)$, that is either in an "on" state (1) or in an "off" state (0) and is characterized by a set of three parametrized values:

$$f_i(t) = \begin{cases} 1 \text{ if } t \in [nx_i + y_i, nx_i + z_i] \text{ for } n \geq 0, \\ 0 \quad\quad\quad\quad\quad \text{else}, \end{cases}$$

$$\text{and } y_i + z_i \leq x_i, \tag{1}$$

where $x_i$ is the period of the signal, $y_i$ is the time within the period when the signal enters the "on" state and $z_i$ is the time within the period when the signal enters the "off" state. A signal can then be easily described by two main features: its period now represented by $x_i$ and the fraction of its period spent pulsing (in the "on" state), which we refer to as a signal's "absolute duty cycle", $d$, and can be expressed as:

$$d = \frac{(z_i - y_i)}{x_i}. \tag{2}$$

This allows our Firefly ETI Model to more efficiently handle the magnitude differences in pulsar periods. It reduces the computational resources required when working with many background pulsars, and more easily calculate all possible time frames when the signals are in the same state, while accounting for phase shift.

The similarity value is proportional to the longest time window where the two signals are identical, calculated over all phase offsets. To determine the similarity value between the artificial signal, $f_1(t)$, and a background pulsar signal, $f_2(t)$, a matching function, $M(t)$, identifies when both signals are in the same state ("on" or "off") and is defined by:

$$M(t) = \begin{cases} 1 & \text{if } f_1(t) = f_2(t), \\ 0 & \text{else} \end{cases} \quad (3)$$

To account for phase shifts and phase offsets—and to capture all time frames when the signal pair is in the same state—we generate multiple permutations of the signal pair (each a specific phase offset/shift), identify the largest time frame of overlap across all permutations, $L$, and define the similarity value, $v$, as the fraction of the longest period spent in the same state at best alignment:

$$v = \begin{cases} \frac{1}{L} & \text{if } f_1(t) = f_2(t) \; \forall \; t, \\ \frac{L}{T} & \text{else} \end{cases} \quad (4)$$

This normalization captures overlap on appropriate time scales and avoids bias toward shorter sequences. For example, a pair with periods $100$ s and $60$ s, and $L = 30$ s yields $v = 0.30$; so does a pair with periods $10$ s and $6$ s, and $L = 3$ s, indicating equal similarity. If the two signals are identical in both absolute duty cycle and period, it yields $v = 1$. We use the maximum overlap because it cannot be guaranteed that the signals will be observed when they are most dissimilar— and because the periods of pulsars are generally not rational multiples of one another, so over long time scales all phase offsets are equally frequent. Minimizing $v$ under this criterion maximizes dissimilarity between the artificial signal and background pulsars while still accounting for the periods when they are most similar. Altogether, this ensures that every signal-pair is weighted equally when averaged later and enables fair comparison across every pairing of the artificial signal with each background pulsar.

The similarity values, $v_i$, are summed and divided by the total number of background pulsars, $n$, to produce an averaged similarity score, $s$, between 0 and 1, where 0 indicates absolute distinctiveness and 1 represents a signal that is identical to every pulsar in the background:

$$s = \frac{\sum v}{n}. \qquad (5)$$

This averaging ensures that the model evaluates how changes to the artificial signal affect its overall similarity to all background signals, allowing a structural modification to be chosen even if it increases similarity to some backgrounds, provided it produces a greater overall decrease in similarity to others.

As a model of ETI, the production of artificial signals with radio pulse luminosities like those of pulsars would require a significant energetic investment. To parametrize the potential energetic costs of creating such signal, $f_i(t)$, we assign an energy score, $e$, to the artificial signal, corresponding to the artificial signal's absolute duty cycle. By minimizing this value, an effective constraint can be placed on the length of a pulse as well as the frequency of pulses, thereby constraining the energetic cost to produce the artificial signal. This energy score is calculated as:

$$e = \frac{z_i - y_i}{x_i}, \qquad (6)$$

where $z_i - y_i$ represents the duration of the pulse sequence above the flux cutoff and its absolute duty cycle and $x_i$ is the period of the signal. This results in an energy score between 0 and 1, where 0 represents a signal that never pulses and 1 represents a signal that is constantly "on". The mathematical form, and function of this term, is identical to the predation risk value in the firefly algorithm of Nguyen et al. (2022). Here, we treat energy constraints as the primary selection pressure shaping ETI signal structure—especially for communication over galactic distances—but we also acknowledge that exposure risk (e.g., attracting potentially predatory or strategically adverse civilizations) could select for "radio-quiet" behavior (Haqq-Misra et al. 2013).

Increasing the signal's energy score, and thereby its absolute duty cycle, leads the structure of an artificial signal to become increasingly distinct from that of the typical pulsar; however, a longer absolute duty cycle is also expected to be accompanied by an increase in energetic cost. This trade-off is captured by assigning weights to the similarity and energy scores when calculating the cost of an artificial signal, defined as:

$$C(\%) = 100 \cdot (w_s s + (1 - w_s)e) \qquad (7)$$

where $s$ is the similarity score, $w_s$ its assigned weight, $e$ is the energy score, and $(1 - w_s)$ its associated weight. The similarity weight, $w_s$, can be viewed as the relative adaptive pressure that balances optimizing for similarity against minimizing energy cost. This equation is almost identical to the one used in Nguyen et al. (2022) to model the evolution of distinct firefly flash sequences, though in our model, weights are bound between 0 and 1 while their algorithm's weights were unconstrained. Although the unconstrained and [0,1]-bounded schemes are functionally equivalent, we adopt the bounded form to reduce the model to a single parameter, $w_s$, and to improve the interpretability of the similarity–energy trade-off.

To ensure that the structure of the artificial signal lies within the confines of a typical pulsar, we use the normal distribution of periods and absolute duty cycles of the background pulsars to define the parameter space. This enables filtering out signal structures that possess features already distinguishing them as anomalous or atypical in reference to the pulsars, without the need for an evolutionary distinguishability argument. For example, though pulsars have been observed with periods anywhere from 0.0014 s to 23 s, the mean period of a pulsar is 0.839 s. Pulsars towards the lower range of periods (between 1 and 10 ms) are atypical of the average pulsar as they generally occur when old pulsars with long periods and low flux densities begin accreting material from a companion star which increases their rotation rates (McClare et al., 2007). Similarly, long period pulsars are also not characteristic of the pulsar background as they are the dying remnants of pulsars with low flux densities that have lost most of their rotational energy from hundreds of millions of years of emitting magnetic dipole radiation (Condon et al., 2015).

Finally, the firefly-ETI model described above is highly modular and flexible, allowing users to set the pulsars' center frequency, the minimum flux density for an observable pulse, the coordinates and radius of the search area, the weights for similarity and energy in the cost function, and the resolution of the search. Using these inputs, the model scans the defined parameter space at the chosen resolution, calculating the cost of every possible signal structure, and outputs the one with the lowest cost (i.e., the global minimum), therefore producing an artificial sequence of flashes that is evolutionarily optimized relative to the surrounding background sources.

# Results

**The Firefly-ETI Model**

We consider a background of 158 pulsars, obtained by setting the Earth as the center of a 5 kpc search area and considering a flux cutoff of 1 mJy at 1400 MHz. We subsequently generated artificial signals with a cost search space resolution of 100x100 and exploring the impact of different $w_s$ values (0.0, 0.2, 0.5, 0.8, and 1.0), corresponding to different trade-offs between dissimilarity and energetic cost. The resolution was restricted to 100x100 for the search space to reduce the computational burden of running the model while still maintaining a level of detail sufficient to producing an accurate artificial signal (See SI, Section III.).

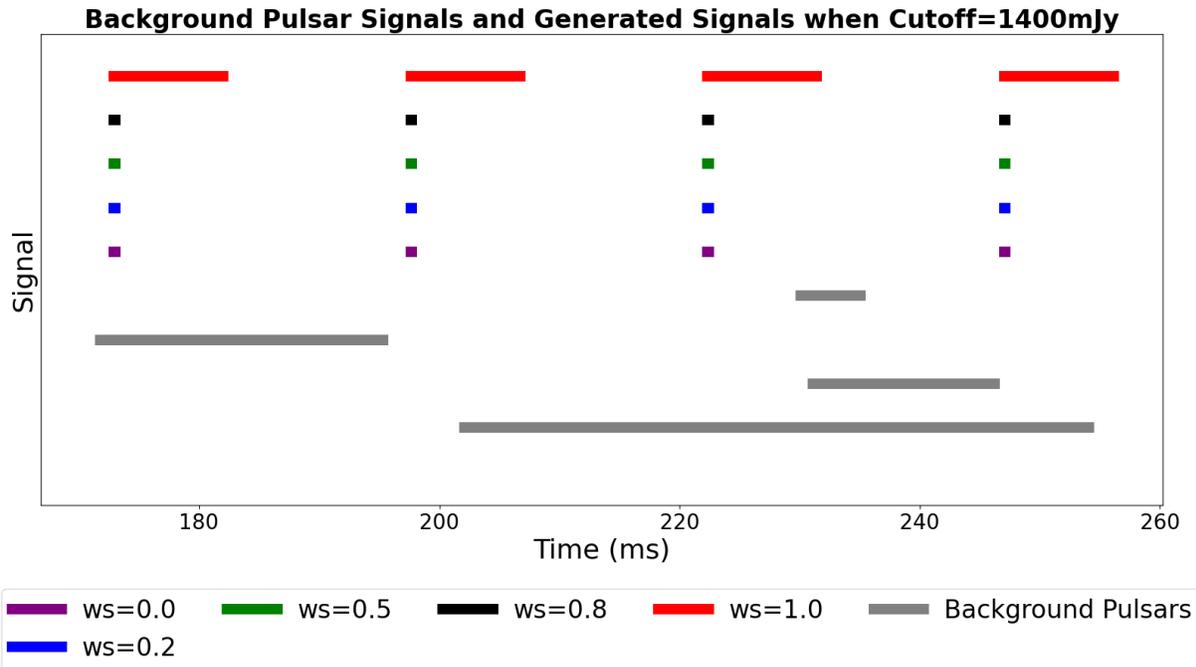

**Figure 2:** Visualization of artificial signals (colored) and example set of representative background pulsars (in grey). The rectangles depict the length of the absolute duty cycle of each signal. The similarity and energy weights have a large influence on the structure of the artificial signal produced, with the absolute duty cycle decreasing by a factor of 92 when $w_s$ decreases from 1.0 to 0.8.

The weight assigned to the similarity score, $w_s$, has a substantial impact on the structure of the artificial signal. As shown in **Fig. 2**, changing $w_s$ from 1.0 to 0.8 decreases the absolute duty cycle of the signal by a factor of 92. However, when energy is considered as a constraint ($w_s \neq 1$), the structure of the artificial signal is identical, regardless of $w_s$ value used. This demonstrates that there are only two distinct solutions for a pulse signal which demonstrates evolved distinguishability from its background depending on whether energy is a constraint to the ETI producing it.

We also applied the firefly-ETI model's cost function to each background pulsar individually, assigning a cost value that reflects its relationship to the rest of the pulsar background (i.e., excluding both the artificial signal and the pulsar itself) across all specified $w_s$ values. The mean and standard deviation of background-pulsar costs were calculated for each $w_s$ value, and a standard score was then computed for each signal pair to show how many standard deviations its cost lies from the mean. This allows confirming that the model produces the lowest cost signal given the parameter space boundaries, resolution, and assigned weight as well as allowing for the potential identification of current signals that may be of ETI origin. We observed that most of the pulsar population have costs that are much higher than that of the optimized artificial signal. More specifically, for $w_s = 0.0$, the artificial signal cost was lower than 84.30 % of the background, and for $w_s = 1.0$ it was lower than 99.78 % of the pulsar background, as illustrated in **Fig. 3** (See SI, Section IV for more details). This suggests that pulsars in the background population do not exhibit evidence of their signals being artificial in origin (as signified by evolutionary optimization). Only two pulsars, J1748-2446C and J1824-2452A, were identified with costs that are equal to or lower than an evolved, artificial signal's cost when energy constraints are not a limiting factor ($w_s = 1.0$). However, both pulsars have periods between 1 and 10 ms, suggesting that these are likely millisecond pulsars that formed naturally through accretion, which would explain their highly atypical periods and signal structure. Note that the artificial signal is constrained by how fine-grained the search space is, and its cost could theoretically be further reduced with access to greater computational power. This would decrease even more the number of pulsars in the population with costs at or below that of the artificial signals, possibly including these two outliers. To further investigate the nature of these two signals, the boundaries of the parameter space were augmented to search specifically within a period range typical for millisecond pulsars (1 ms to 10 ms); an artificial signal was generated with a cost of 0.979 %, much lower than that of the two anomalous pulsars (2.65 % and 1.03 %).

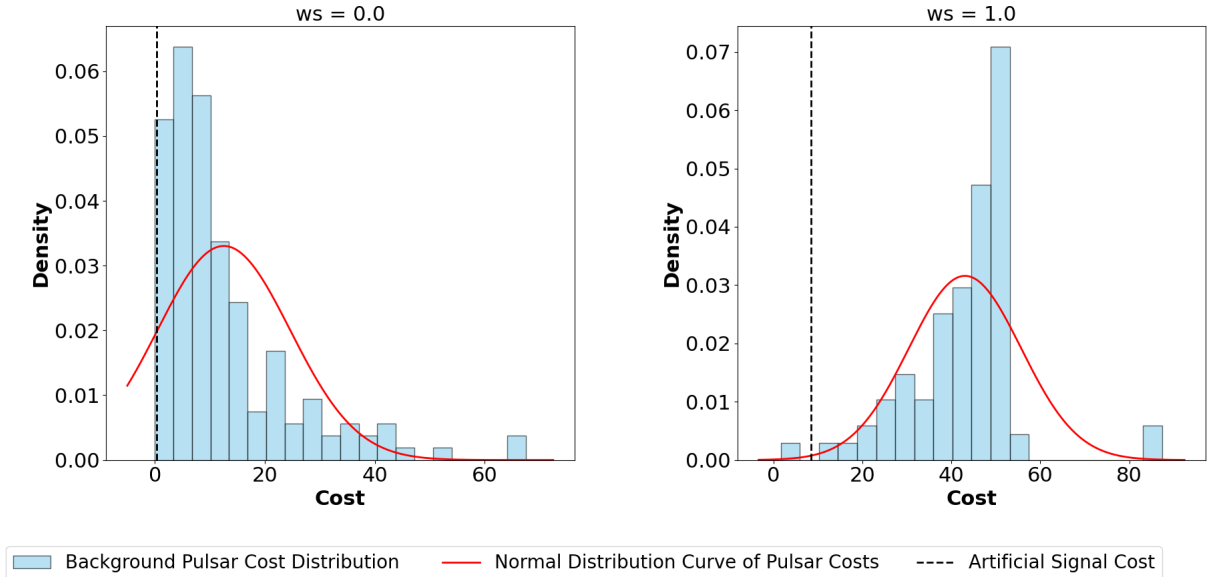

**Figure 3:** Normal distribution of the assigned costs of the 158 background pulsars within 5 kpc of Earth, compared to the cost of the optimized artificial signal produced by the model when $w_s = 0.0$ **(left)** and $w_s = 1.0$ **(right)**. Most pulsars within the background have costs that are above that of the artificial signal, with 84.30 % of pulsars displaying costs above that of the artificial signal when $w_s = 0.0$, compared to 99.69 % when $w_s = 1.0$.

Concerning the distribution of signal costs, it is strongly influenced by the weights assigned to the similarity and energy scores, as seen in **Fig. 4**. However, regardless of the value of $w_s$, the lowest-cost artificial signals consistently appear near the boundaries of the search space, which coincide with the extremes of the pulsar period distribution. This outcome is not unexpected, since signals at either extreme are relatively rare in the pulsar background or stand out as more distinct. This is a feature of the relatively low dimensionality of a binary code, but we might expect more complex landscapes for more complex languages. Our case study may be biased toward favoring boundary solutions, rather than identifying potentially distinct signals within the more densely populated interior of the distribution.

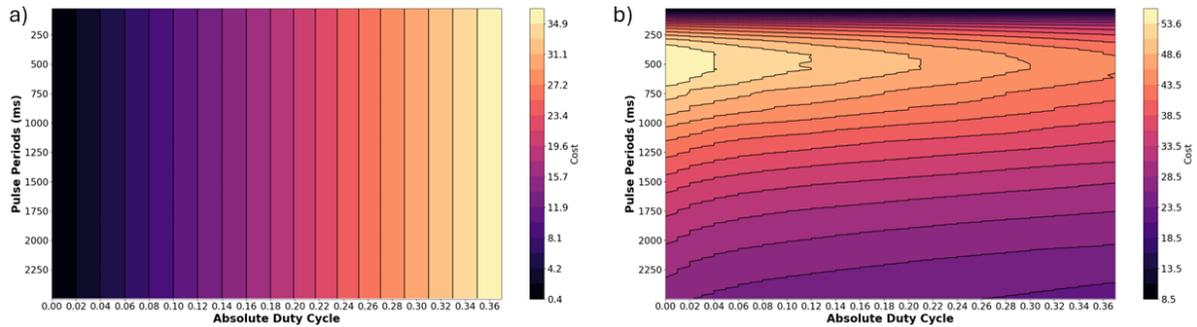

**Figure 4:** Cost landscape of artificial signals defined by the period and absolute duty cycle of pulse sequences, for a similarity weight of **a)** $w_s = 0.0$ (i.e. energy limitation as the sole factor)—signals with lower costs congregate around the lowest absolute duty cycle boundary, with a uniform increase in artificial cost as the absolute duty cycle increases—and **b)** $w_s = 1.0$ (i.e., no energy limitations)—signals with lower costs congregate around the period boundaries, particularly when the period of the artificial signal is shorter than 272 ms.

In the cost landscape where energy constraints are strong, such as in **Fig. 4a**, a uniform, gradual slope is observed moving from an area of low cost when the absolute duty cycle is near zero towards an area of high cost as the absolute duty cycle increases. The dominance of this slope in the topography of this cost landscape reveals that when $w_s < 1.0$, the signal's absolute duty cycle becomes the dominant signal feature affecting the artificial signal's cost and thereby its final structure. It is worth noting that the period of the artificial signal still has a minor influence on its cost and will produce a steeper gradient the shorter the period becomes (< 272 ms).

When energy consumption is a significant factor, the lowest cost artificial signal adopts a structure reducing the amount of time spent "on" in a tradeoff with dissimilarity. Consequently, the trend towards lower cost signals comprised of short periods or small absolute duty cycles is expected. When energy is not a constraint ($w_s = 1.0$), the topography of the cost landscape does not exhibit this slope, see **Fig. 4b**. Instead, there is a gradual uniform slope from an area of low cost where the period is long, to a hill of high cost between a period of 272 ms and 766 ms. The cost then falls sharply as the period decreases further below 272 ms, creating a steep cliff of low cost. The strong correlation between the periods and the costs of these artificial signals is expected under the assumption that access to energy is not a constraining factor. In this case, adjusting the period is the dominant way in which an artificial signal could distinguish itself from the background. Furthermore, as predicted, with no energy constraint, the more dissimilar artificial signals congregate around the upper bounds of the absolute duty cycle. In the case of this background, the most dissimilar artificial signal has an absolute duty cycle close to 0.36 on this upper bound, and a period that is just under 25 ms and firmly within the steep cliff of low cost.

The firefly–ETI model identifies the signal structure that is best optimized to stand out from the pulsar background, while remaining within a parameter space that is realistic for a typical pulsar and consistent with the energy limitations of the signal producer. We observe two distinct solutions for the structure of a pulse signal demonstrating evolved distinguishability from the pulsar population within 5 kpc of the Earth with a mean flux cutoff greater than 1 mJy, differentiated by whether energy is a limitation to the artificial signal creator.

The firefly-ETI model exhibits significant departure from some of the computational strategies in Nguyen et al. (2022). Most notably, the fitness landscape of the possible pulse sequences is simpler than in the case of firefly flashes, as highlighted in **Figure 3**. This allows us to compute all the possible signals within the parameter space constrained by the distribution of background pulsars and search for a global minimum in the associated costs. This is significantly different from the evolutionary, "signal mutation"-based algorithm they used to find optimally distinguishable sequences, but the implications and results remain the same. While the evolutionary relatedness between humans and fireflies is undoubtedly closer than that with alien species, this approach allows for a more comprehensive exploration of communication space in SETI and METI efforts, potentially increasing the likelihood of successful detection. Moreover, the use of burst-based and patterned sequences for communication is not unique to fireflies but has convergently evolved across multiple lineages on Earth and can be seen in birdsongs, woodpecker knock drumming, frog calls, eel electroshocks (e.g., Garcia et al. 2020), and human-made morse codes. This lends credence to the notion that the firefly-inspired method may represent a general solution to the communication challenge, potentially applicable beyond Earth's biosphere.

**Speculations on Firefly-ETI**

The generation and interpretation of an artificial signal depends on the pulsars that are included in the background population. This has several implications regarding the level of intelligence necessary to engineer such a signal. For instance, for a pulsar to be visible from Earth, the beams of radiation emitted from its poles must intersect our line of sight and remain unobstructed by interstellar material or other astrophysical objects. Thus, any communicative intelligence driven by the principles of our model would need to have an intimate understanding of which pulsars would be visible to its target. That is, the ETI civilization would necessarily need to be advanced enough to have a 'theory of mind' for other technological life that could observe the signal (e.g., us), and sufficient knowledge of physics to be able to model the pulsar environment of Earth. The ability to perceive the mental state of others has been observed arising in humans between individuals similar, as well as substantially different, cultures (Selcuk et al., 2022) and is suggested to be present in several other animal species on Earth (e.g., the Corvidae family, as discussed in Chopra, 2022). Consequently, if this ability is widely shared across species and social systems on Earth, it is reasonable to infer that an ETI—particularly one engaged in the search of other intelligences—may possess a functionally analogous form of theory of mind.

The model's artificial sequences are influenced by the weight assigned to the similarity score, and consequently the energetic cost, raising critical questions about the prescription of these parameters. The determination of this weight would likely be contingent upon the method of pulse production, the associated energy requirement, and the available energy resources of the intelligence in question. It has often been posited that access to energy is closely tied to technological advancement, a sentiment that is captured in the Kardashev Scale. Kardashev (1964) proposed that a civilization's energy consumption was closely related to its technological development, based on the realization that the energy consumption of humanity would increase by approximately 3-4 % over the next 60 years. He classified civilizations into 3 types: Type 1 civilizations are at a technological level close to current day human civilization with energy consumption around $4 \times 10^{19}$ ergs/s, Type 2 civilizations are more technologically advanced and are able to harness all the energy radiated by their own star with an energy consumption around $4 \times 10^{33}$ ergs/s, and Type 3 civilizations are significantly more advanced and able to leverage the energy of the entire galaxy with a consumption around $4 \times 10^{44}$ ergs/s.

We can consider our results in the context of the Kardashev-analogous, energy-consumption scale an ETI would need to reach to produce an artificial firefly-like pulsing signal of equal luminosity to a pulsar. To approximate the energetic cost of creating a pulse sequence like the ones outlined by the model, one must consider the period of a pulsar increases over time by losing some of its rotational kinetic energy as magnetic dipole radiation. Assuming this is the only energy loss mechanism, the energy released as radiation by a pulsar must be equal to its spin down luminosity. Using the Crab pulsar from the Crab nebula as a proxy for this calculation, the amount of energy needed to produce the magnetic dipole radiation of a pulsar would be $10^{38}$ergs/s, thus requiring $\sim 10^{30}$ergs/s to produce the resulting average radio pulse luminosity, as calculated using the following equation:

$$-E = \frac{4\pi^2 IP}{p^3}, \qquad (8)$$

where $E$ is the spin down luminosity, $I$ is the pulsar's moment of inertia, $P$ represents the period derivative, and $p$ is the pulse period (Condon et al., 2015). Given these quantitative approximations, any extraterrestrial intelligence using this method to broadcast their presence is likely to be far more advanced than the current level of technology on Earth. Only Type 2 and Type 3 civilizations could have sufficient energy to produce these signals, and so without concern for their energy requirements. Therefore, we would expect them to produce the types of artificial signal structures exhibited when $w_s$ = 1.0. However, since the absolute duty cycles of these signals can vary between milliseconds per period to seconds, a less technologically advanced ETI (though still more advanced than humanity) could still optimize for energy consumption minimization and produce these types of signals with structures reminiscent of those seen when $w_s$ < 1.0. Thus, under the assumptions that pulse signal production is energy intensive and access to energy directly correlates to technological advancement, the introduction of the energy cost into the firefly-ETI model may provide information on the technological level of the signal producer relative to our own. Altogether, this suggests that more advanced civilizations (higher on the Kardashev Scale) might produce signals with significantly higher similarity weights and thus lower energy score weights.

Note that in the current version of the model, we do not account for the fact that the period of a pulsar decreases over time as some of its rotational kinetic energy is converted into magnetic dipole radiation and emitted. The resulting change in the structure of the pulsars over time is likely to affect the recognizability of the artificial signal depending on the distance travelled to the target. However, the change in a pulsar's period over time is typically well known and can be accounted for. Thus, we believe that not including the effects of time on the pulsar periods does not seriously detract from the findings of this model.

# Discussion

On Earth, the first communication signal of an individual/species typically establishes their presence first and foremost (akin to "I am here") before trying to pass any further meaning. For instance, evidence suggests that identity-signaling, or "contact calls" are a fundamental component of almost all species, independent of social structure (e.g., Kondo & Watanabe, 2009; Berg et al., 2011; McComb et al., 2003; Sayigh,1999). This universality is also reflected by the importance of human "phatic" communication (Zuckerman, 2011). Because this elementary presence-establishing signal is ubiquitous among species on Earth, it is likely to occupy a privileged role in communication and, by extension, generalize beyond our planet. However, very few signals carry such obvious semantic content.

Within SETI, two principal paradigms have traditionally guided approaches to detecting extraterrestrial intelligence (Tarter, 2001; Vakoch, 2014). Proponents of the *detection-first* strategy believe that a focus on identifying signatures of ETI without necessarily understanding their meaning or function will be the most efficient means of finding ETI (Doyle et al., 2011). On the other hand, supporters of the *decoding-first* strategy argue that prioritizing the analysis of incoming signals for meaning and messages will provide a more dependable and definitive means of identifying ETI (Matessa et al., 2022). Artificial signals generated by the Firefly ETI model are intended to demonstrate how ETI can be detected via their evolved distinctiveness relative to an abiotic background, which allows inferring the presence of an intelligent signal without presupposing or depending on semantic content; this underscores the advantage of a detection-first strategy over a decoding-first approach. SETI efforts are more likely to be successful when incorporating the search for signals that manifest structure-driven principles of life through a more global understanding of communicative strategies, rather than focusing on human-centric and complex semantic content of observations that do not demonstrate any structural differentiation.

The firefly's distinct flash sequence provides a means of identifying members of the same species for mating. But importantly this meaning behind the signaling need not be known to identify the distinct patterns different species generate. Our focus on the firefly model is intended to provide a quantitative thought experiment to probe how searching for an evolved signal could allow detecting the presence of alien intelligence. It is worth noting that we use "evolved" rather than "engineered' when referring to these maximally distinct signals as a deliberate conceptual reframing–it forces reducing anthropocentric bias, recognizing that evolution encompasses a broader and more universal framework than the human-centric notion of engineering (Jacob 1977). The key features of our model, which should be generalizable to other biological species models and ideally even alien species, is that we assume the structure properties of communication will be recognizable first (and inference of meaning will come later) and that these are to be understood as structural properties that are the product of an evolutionary process.

The integration of insights from non-human communication research into SETI and METI efforts presents a novel approach to understand the potential varieties of interstellar extraterrestrial communication focusing on the structure of communication and not its meaning. The toy model presented here is not intended as a fully realistic or empirically predictive framework; rather, it presents a thought experiment imagining what kinds of alien intelligences may be possible and illustrates the potential to develop new strategies for the search for extraterrestrial intelligence (ETI) by extending communication models beyond human paradigms to include non-human species. Our aim is to motivate approaches that reduce anthropocentric bias by drawing on different communicative strategies observed within Earth's biosphere. Such perspectives broaden the range of ETI forms we can consider and leverage a more comprehensive understanding of life on Earth to better conceptualize the possible modes of extraterrestrial communication (Doyle, 2011). Broadening the foundations of our communication model, by drawing systematically from diverse taxa and modalities, would yield a more faithful representation of Earth's biocommunication and increase the likelihood of success, with less anthropocentric searches, and more insights into deeper universalities of communication between species. To achieve this shift, the international community should encourage cross-disciplinary research projects to establish transparent criteria for selecting and optimizing model species diversity in SETI research, thereby mitigating selection bias and promotive comparative, theory-driven work.

# Acknowledgements

We thank Nicholas Barendregt for assisting us to understand the Firefly-Inspired Vocabulary Generator code and how that could be translated into our model and Peyton Idleman for her help drafting the mathematical representations of our model. This work was supported by a Schmidt Science Polymath Fellowship awarded to S.I.W and to O.P.

# Supplemental Information

**Section I: Firefly Vocabulary Calculations**

In the Firefly Vocabulary Algorithm developed in (Nguyen et al., 2022), a flash sequence of length $L$ is defined as a binary sequence of 1s and 0s where an instance of 1 would represent that the sequence is "on or "flashing" and 0 depicting that the sequence is "off or not flashing." Such that:

$$(a_n)_{n=1}^{L}, a_n \in \{0,1\} \qquad (S1)$$

The cost function which drives the creation of flash sequences in the algorithm is influenced by two features that are believed to strongly shape the influence the flash sequences fireflies use to communicate. These two features being their distinguishability to other species' flash patterns, and predation risk (the likelihood that the flash pattern attracts unwanted attention from a predator):

$$C((a_n)_{n=1}^L) = w_s s + w_p p \tag{S2}$$

The average similarity between the flash sequence and all other species' sequences, $s$, and the predation risk of the flash sequence, $p$, are affected by weights, $w_s$, and $w_p$. The relative values of these weights can be augmented to adjust the ratio of average similarity influence to predation risk influence, $\frac{w_s}{w_p}$, to reflect environmental conditions and thereby affect the features of the evolved flash sequence.

The similarity between two flash sequences, $(a_m)_{m=1}^L)$ and $(a_n)_{n=1}^L$ is expressed as the length of the longest common substring under cyclic permutation:

$$LCS((a_m)_{m=1}^L, (a_n)_{n=1}^L) \tag{S3}$$

This represents the maximal possible overlap between the sequences that accounts for all cyclical permutations since there is no guarantee that two firefly species will begin flashing at the same time or a constant offset.

$$LCS((a_m)_{m=1}^L, (a_n)_{n=1}^L) = \max_{0 \leq k \leq L-1} f((a_m)_{m=1}^L - (a_n \to a_{n+k(mod\ L)})_{n=1}^L) \tag{S4}$$

In the equation above, $a_n \to a_{n+k}$ represents moving the $nth$ term of the flash sequence $k$ places to account for cyclical permutations and the function $f(a_n)_{n=1}^L$ computes the length of the longest consecutive subsequence of zeros in the sequence. The averaged similarity value, $s$, is found by averaging the $LCS$ of all sequence pairs and normalizing by the length of the binary flash sequences so that s is a value between 0 and 1.

The predation risk, $p$, is expressed as the fraction of bits in the "on state" of the entire sequence length that will also result in a value between 0 and 1.

$$p = \sum_{n=1}^L \frac{a_n}{L} \tag{S5}$$

Consequently, a predation risk, *p*, of 1 would indicate high predation risk and a predation risk, *p*, of 0 would indicate no predation risk. This is done as it is presumed that flashing more frequently or for longer durations is more likely to alert a predator to the flash sequence creator's presence and location thus increasing the risk of being preyed upon.

## Section II: Firefly ETI Model Pulsar Data Set & User Defined Parameters

The pulsar data that the Firefly ETI Model utilizes to generate the pulsar background is retrieved from the Australia National Telescope Facility (ATNF) pulsar database (Manchester et al., 2005) utilizing an interface developed by Pitkin (Pitkin, 2018). This interface functions as a python module called *psrqpy* which allows users to query the ATNF pulsar catalogue as a script instead of having to download and analyze the text tables generated by the ATNF's web interface to access the pulsar information. The ATNF has information on 3724 pulsars, which can be filtered according to several parameters in the web interface and thus by extension by the *psrqpy* python module as well. For the purposes of the Firefly ETI Model, the parameters used to filter the pulsar data were: galactic position, the radio frequency of interest, and the minimum mean flux density of the pulsars.

To determine the appropriate pulsar background to use to generate the evolved artificial signal, the Firefly ETI Model prompts the user to enter a central location using galactic coordinates in kiloparsecs where Earth is defined at the origin (0,0,0), as well as a search area radius (in kpc). The model will then filter through the ATNF catalogue and select only the pulsars with galactic coordinates that fall within the search radius from the central location the user defined.

Pulsars emit over a large range of radio and x-ray frequencies, however differences in where within the pulsars these emissions occur, and how different frequencies are affected when travelling through the interstellar medium can affect the brightness and width of the pulses when received by our telescopes. Thus, the model will prompt the user to define the radio frequency of interest in order to focus on the pulsar characteristics at the specified frequency.

Mean flux density describes the average amount of energy per unit area per unit frequency per unit time over the duration of a pulsar's rotation period, and as such can be used to quantify the amount of energy emitted by a pulsar or its "brightness" at specific frequencies. As such the Firefly ETI model will use the minimum mean flux density provided by the user to filter out pulsars from the ATNF database that have mean flux density's lower than the given value and as such would be considered "too dim" at the selected search frequency.

**Section III: Firefly ETI Model Parameter Space Definition & Resolution**

In the Firefly ETI Model, the boundaries of the signal structure parameter space and chosen resolution affect the number and specific signal structures explored by the model as well as the structure and cost of the lowest cost artificial signal ultimately produced by the model. The parameter space boundaries are defined by the normal distribution of the periods and absolute duty cycles of the pulsar background being investigated. As such the upper limit on the period and absolute duty cycle is placed 2 standard deviations above the mean period and mean absolute duty cycle respectively, $(\mu + 2\sigma)$. Similarly, the lower limit of the period and absolute duty cycle of potential artificial signals is placed 2 standard deviations below the mean period and mean absolute duty cycle $(\mu - 2\sigma)$. The resolution provided by the user then defines the number of points between the upper and lower bounds that will be investigated as potential features of the artificial signal. In this paper, we used a resolution of 100 x 100 and thus explored 10,000 combinations of the 100 periods and 100 absolute duty cycles within the normal distribution of the pulsar background for the artificial signal structure with the lowest cost value accounting for the $w_s$ value assigned.

The effect of the chosen resolution on the specific signal structures explored by the Firefly ETI Model is noticeable; however, these effects do not appear to significantly alter the cost or the structure of the artificial signal produced by the model. For example, changing the resolution from 50x50 to 100x100 only decreased the cost value of the produced artificial signal by approximately 0.05%, and the signal structure occupied roughly the same location in the parameter space. This makes sense as the parameter space regarding the actual structures of pulsars is not incredibly dense and as such, increasing the resolution of the investigation does not yield significantly different results.

The boundaries of this parameter space on the other hand have a pronounced effect on the signal structures explored by the Firefly ETI Model, and consequently the cost of the artificial signal produced. The Firefly ETI Model operates by searching for the artificial signal structure that has the lowest cost value and thus consequently is the most dissimilar from the pulsar background while accounting for any energy requirements as defined by $w_s$ value. Consequently, as the boundaries of the parameter space are expanded further away from that typical of most pulsars, the signal structures that are created at these boundaries become increasingly atypical of pulsars and thus also increasingly distinct from pulsars. Identifying this, the model then produces an artificial signal with a structure towards these boundaries that are inherently atypical of the pulsar population. Thus, changing the boundaries of the parameter space can have a significant effect on the structure and cost of the artificial signal and becomes more notable as the change increases.

**Section IV: Firefly ETI Model Data & Results**

Using the 158 pulsars in the pulsar population within 5kpc of the Earth with mean flux densities greater than 1mJy, and a parameter space resolution of 100x100, the Firefly ETI Model produced a pulse signal that displayed evolved distinguishability from the pulsar background for each assigned $w_s$ value. The signal structure of the artificial signals produced when $w_s = 0.0$, $w_s = 0.2$, $w_s = 0.5$, and $w_s = 0.8$ were all identical having a period of 24.704s and an absolute duty cycle of 0.004. When $w_s = 1.0$, the optimized pulse signal still had a period of 24.704s but instead had an absolute duty cycle of 0.368. The costs of these signals were 0.400 %, 2.139 %, 4.747%, 7.355% and 8.526% respectively.

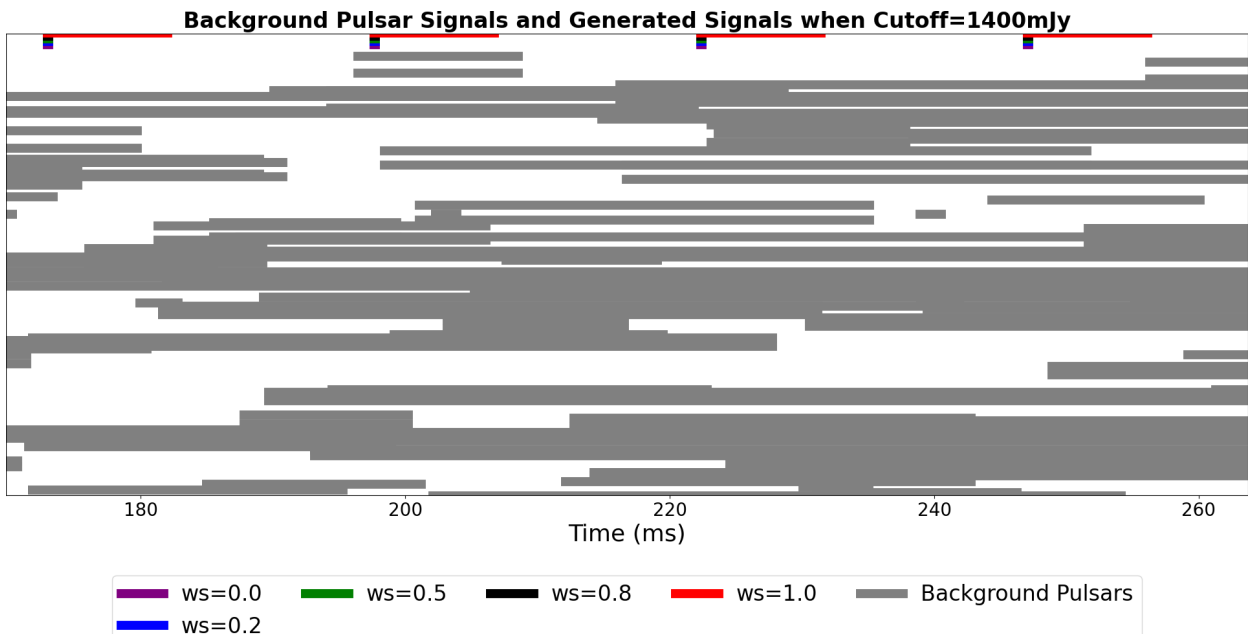

**Figure 5:** Visualization of artificial signals (colored) and the 158 pulsars that make up our target background (in grey). The rectangles depict the length of the absolute duty cycle of each signal. The similarity and energy weights have a large influence on the structure of the artificial signal produced, with the absolute duty cycle decreasing by a factor of 92 when $w_s$ decreases from 1.0 to 0.8.

When these costs were compared to the normal distribution of pulsar costs attained by applying the Firefly ETI Model's cost function to each pulsar of the background at each $w_s$ value, it was discovered that at $w_s = 0.0$ the artificial pulse signal had a cost lower than 84.30 % of background pulsar costs, at $w_s = 0.2$ the artificial pulse signal was lower than 96.50% of the background distribution, at $w_s = 0.5$ it was lower than 99.94% of the background, at $w_s = 0.8$, it was less than 99.87% of background pulsar costs, and when $w_s = 1.0$ the cost of the artificial pulse signal's structure was less than 99.69% of the pulsar background. The reason the artificial signal structure at $w_s = 1.0$ was not as distinguished from the background pulsar costs as the artificial signal structure under the $w_s = 0.8$ conditions is a result of the inclusion of the two millisecond pulsars in the background. The structure of these pulsars is naturally atypical of a pulsar and as such, when energy constraints are not considered, these pulsars had costs lower than that of the artificial signal produced at $w_s = 1.0$ within the normal distribution of the pulsar background's periods and absolute duty cycles. This affected the distribution of background pulsar costs, skewing the mean cost lower than it would have been otherwise, thus reducing the percentage of the normal distribution of costs that the artificial signal's cost was lower than.

Earth". *Science.* (Vol. 199, Issue 4327, pp.377-388).
https://doi.org/10.1126/science.199.4327.377

Tarter, J. (2001). "The Search for Extraterrestrial Intelligence (SETI)". *Annual Review of Astronomy and Astrophysics.* (Vol. 39, pp. 511 548).
https://doi.org/10.1146/annurev.astro.39.1.511

Ten Cate, C., (2004). "Birdsong and evolution*".. Nature's Music*. (Chapter 10, pp. 296–317).
https://doi.org/10.1016/B978-012473070-0/50013-X

The Staff at the National Astronomy and Ionosphere Center. (1975). "The Arecibo message of November, 1974". *Icarus*. (Vol. 26, Issue 4, pp. 462-466). https://doi.org/10.1016/0019-1035(75)90116-5

Vakoch, D. A. (Ed.). (2014). "Archaeology, Anthropology, and Interstellar Communication". NASA History Series. Washington, DC: National Aeronautics and Space Administration.

Wang, Z., Wang, J., Wang, N., Dai, S., & Xie, J. (2023). "Study of pulsar flux density and its variability with Parkes data archive". *Monthly Notices of the Royal Astronomical Society*, (Vol. 520, Issue 1, pp. 1311–1323). https://doi.org/10.1093/mnras/stad199

Woods Jr., W. A., Hendrickson, H., Mason, J., Lewis, S. M. (2007). "Energy and Predation Costs of Firefly Courtship Signals". *The American Naturalist.*(Vol. 170, No.5, pp.702-708). https://doi.org/10.1086/521964

Worden, S. P., Drew, J., Siemion, A., Werthimer, D., DeBoer, D., Croft, S., MacMahon, D., Lebofsky, M., Isaacson, H., Hickish, J., Price, D., Gajjar, V., & Wright, J. T. (2017). "Breakthrough Listen – A new search for life in the universe". *Acta Astronautica*. (Vol. 139, pp.98–101). https://doi.org/10.1016/j.actaastro.2017.06.008

Wright, J. & The Conversation. (2017). " Voyager Golden Record 40 Years Later: Real31